\documentclass[pdflatex,sn-nature]{sn-jnl}% Style for submissions to Nature Portfolio journals
\usepackage{geometry}
\usepackage{graphicx}%
\usepackage{multirow}%
\usepackage{amsmath,amssymb,amsfonts}%
\usepackage{amsthm}%
\usepackage[title]{appendix}%
\usepackage{xcolor}%
\usepackage{textcomp}%
\usepackage{manyfoot}%
\usepackage{booktabs}%
\usepackage{siunitx}
\usepackage{algorithm}%
\usepackage{algorithmicx}%
\usepackage{algpseudocode}%
\usepackage{listings}%
\usepackage{placeins}

\begin{document}

\title[Surface impedance inference via neural fields and sparse acoustic data obtained by a compact array]{Surface impedance inference via neural fields with sparse acoustic data obtained by a compact array}

\author*[1]{\fnm{Yuanxin} \sur{Xia}}\email{yuxia@dtu.dk}

\author[1]{\fnm{Xinyan} \sur{Li}}\email{lxy19181100@outlook.com}
\author[1]{\fnm{Matteo} \sur{Calaf\`a}}\email{matcal@dtu.dk}
\author[2]{\fnm{Allan P.} \sur{Engsig-Karup}}\email{apek@dtu.dk}
\author*[1]{\fnm{Cheol-Ho} \sur{Jeong}}\email{chje@dtu.dk}

\affil*[1]{\orgdiv{Acoustic Technology, Department of Electrical and Photonics Engineering}, \orgname{Technical university of Denmark}, \orgaddress{\street{Ørsteds Plads, 352}, \city{Kongens Lyngby}, \postcode{2800}, \country{Denmark}}}

\affil[2]{\orgdiv{Department of Applied Mathematics and Computer Science}, \orgname{Technical university of Denmark}, \orgaddress{\street{Asmussens Allé, 303B}, \city{Kongens Lyngby}, \postcode{2800}, \country{Denmark}}}

\abstract{Standardized laboratory characterizations for absorbing materials rely on idealized sound field assumptions, which deviate largely from real-life conditions. Consequently, \emph{in-situ} acoustic characterization has become essential for accurate diagnosis and virtual prototyping. We propose a physics-informed neural field that reconstructs local, near-surface broadband sound fields from sparse pressure samples to directly infer complex surface impedance. A parallel, multi-frequency architecture enables a broadband impedance retrieval within runtimes on the order of seconds to minutes. To validate the method, we developed a compact microphone array with low hardware complexity. Numerical verifications and laboratory experiments demonstrate accurate impedance retrieval with a small number of sensors under realistic conditions. We further showcase the approach in a vehicle cabin to provide practical guidance on measurement locations that avoid strong interference. Here, we show that this approach offers a robust means of characterizing \emph{in-situ} boundary conditions for architectural and automotive acoustics.
}


\maketitle

\section{Introduction}\label{introduction}
The acoustic behavior of an enclosed space is governed by its boundary conditions, typically characterized by either the absorption coefficient or surface impedance. While real-valued absorption coefficients are widely adopted for their simplicity, their primary limitation lies in their non-uniqueness, as distinct complex impedance values can yield identical absorption coefficients. Consequently, a rigorous physical characterization requires retrieval of complex-valued surface impedance. Furthermore, the acoustic performance of absorbers \emph{in-situ} often deviates from laboratory-controlled results, rendering \emph{in-situ} measurements essential for high-fidelity virtual prototyping in architectural and automotive acoustics~\cite{vorlander2013computer}. Yet standardized tests conducted in anechoic or reverberation rooms rely on idealized assumptions that are seldom realized in practice. While ISO 354 assumes a perfectly diffuse field~\cite{ISO354,zea2023sound,muller2024free}, the impedance tube method is limited to normal incidence~\cite{ISO:10534-2:2023} and is sensitive to mounting of small specimens~\cite{cummings1991impedance,tsay2006influence}.

To address these limitations, several \emph{in-situ} measurement techniques have been suggested over the past few decades to characterize absorbers under actual mounting conditions. Early approaches focused on separating incident and reflected wave components from sound pressure measurements or performing spectral signal processing~\cite{yuzawa1975method,cramond1984reflection,allard1985measurement,mommertz1995angle}. However, these methods usually necessitate free-field conditions and are sensitive to the geometric alignment. A widely adopted approach uses direct particle velocity sensors~\cite{de2003microflown, lanoyeMeasuringFreeField2006}. This approach was solidified by the development of the combined pressure–velocity ($p\text{-}u$) probe, which allows the simultaneous acquisition of sound pressure and particle velocity at a single point near the material surface, allowing for surface impedance estimation with a certain degree of immunity to room reflections~\cite{brandao2011comparison,pedreroAccuracySoundAbsorption2020}. However, the probe is highly sensitive to structure-borne vibrations and temperature fluctuations~\cite{linSituMeasurementAbsorption2016}, and its accuracy degrades when characterizing reflective or near-rigid surfaces, where the particle velocity at the boundary approaches zero. This results in a poor signal-to-noise ratio (SNR) for the velocity sensor, leading to unreliable impedance estimations. Additionally, as a single-point measurement, the technique remains particularly sensitive to local field distortions caused by positioning errors and diffraction effects from finite-sized specimens~\cite{brandaoEstimationMinimizationErrors2012}. To address these uncertainties, recent research has incorporated advanced statistical frameworks, such as Bayesian inference with sequential frequency transfer, to quantify uncertainty for free-field measurements~\cite{eserFreefieldCharacterizationLocally2023}.

As an alternative to the $p\text{-}u$ probe, the local plane wave (LPW) method was introduced~\cite{kuipers2013Measuring}. This approach approximates the complex sound field within a small spatial volume as a superposition of incident and reflected plane waves, enabling the estimation of incident intensity without prior knowledge of the global sound environment~\cite{kuipers2012numerical}. Its accuracy heavily depends on the validity of the local plane wave assumption, so the accuracy can be degraded in the presence of room reflections or near-field conditions~\cite{kuipers2014measuring}, where additional wave components violate the assumed two-plane-wave model.

Similarly, sound field reconstruction (SFR) techniques, such as near-field acoustic holography (NAH) and the equivalent source method (ESM), have emerged as advanced tools~\cite{richard2017estimation,hald2019situ,ottink2016situ}. These methods aim to decompose complex acoustic fields above the material surface and back-propagate the reconstructed field to the surface. They typically rely on dense spatial sampling to resolve the wavefront that single-point sensors fail to capture. While they can achieve high accuracy in controlled anechoic environments, their practical deployment is often constrained by prohibitive hardware requirements, either through large microphone arrays with typically over fifty to hundreds of channels~\cite{ottink2016situ,nolan2025free} or time-consuming spatial scanning, e.g., mechanically or robot-assisted, which are impractical for \emph{in-situ} measurements~\cite{rathsam2015analysis,nolan2020estimation}. Moreover, the robustness of traditional SFR algorithms often relies on free-field assumptions~\cite{nolan2025free} that are violated in reverberant or enclosed environments, leading to significant reconstruction artifacts.

Recently, machine learning techniques have emerged as a powerful alternative for acoustic characterization. Data-driven approaches have demonstrated success in predicting absorption from material microstructure~\cite{yang2022prediction} or estimating impedance from limited pressure measurements~\cite{emmerich2025data}. To quantify parameter uncertainty using the Bayesian method, simulation-based inference has been proposed to infer posterior distributions of the impedance-model parameters by training on extensive precomputed finite-element simulations~\cite{schmid2026situ}. However, these methods typically depend on accurate geometric and physical priors to generate synthetic training data and remain susceptible to the simulation-to-reality gap, which can degrade performance in unseen acoustic environments. To mitigate these challenges, physics-informed neural networks (PINNs)~\cite{raissi2019physics} have been introduced to solve inverse problems by embedding physical laws directly into the learning process~\cite{xia2024neural,schmid2024physics}. While standard PINNs formulations incorporate frequency as an input variable to address the broadband nature of sound, this approach significantly increases the complexity of the optimization landscape~\cite{sandhu2023multi}.

This work proposes a physics-informed machine learning framework for rapid \emph{in-situ} absorption and impedance evaluation. First, we introduce a parallel multi-frequency neural field architecture with sinusoidal encoding to overcome broadband computational bottlenecks. The design enables inference within seconds to minutes on a consumer-grade graphics processing unit (GPU) using only sparse pressure samples. To leverage this algorithmic efficiency, a compact microphone array is developed with a substantially reduced hardware complexity compared with traditional multi-channel arrays~\cite{schmid2024physics,brandao2025sound}. Second, we circumvent the instrumental limitations of velocity sensors by inferring particle velocity via automatic differentiation, thereby avoiding the inherent challenges of particle velocity sensors. Third, unlike LPW or reconstruction approaches that impose restrictive field assumptions, our framework performs physics-informed extrapolation governed by the acoustic equations.

In this work, we validate the framework in three phases, progressing from numerical optimization to experimental deployment and physical mechanism analysis. We begin with parametric simulations to quantify sensor requirements and guide the design of a custom microphone array. The prototype is subsequently validated in both anechoic and reverberant chambers to evaluate its robustness across sound field conditions. Finally, we demonstrate the approach in a virtual vehicle cabin, highlighting its adaptability to highly complex geometries with significant wave interference, and leverage available ground-truth data to correlate reconstruction fidelity with local sound field complexity.

\section{Results}\label{results}
%-------------------------------%
\subsection{Impedance prediction via parallel networks from sparse data}\label{sparse_data}
\begin{figure*}[!ht]
    \centering
\includegraphics[width=\linewidth]{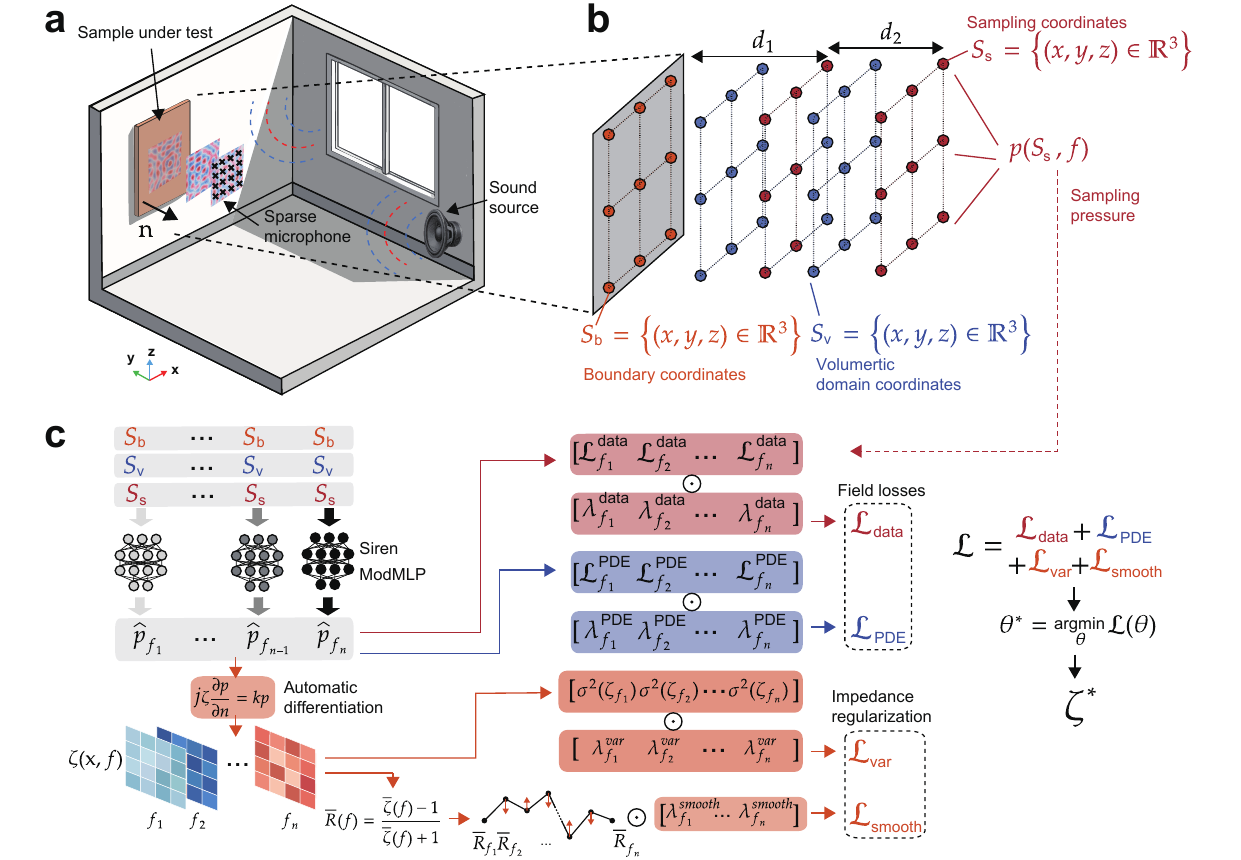}
\caption{\textbf{Framework for surface impedance inference using parallel neural fields.} \textbf{a}, Schematic concept: sparse pressure measurements near a material surface serve as input for characterizing the boundary condition. \textbf{b}, Spatial definition of the computational domain: boundary points ($S_b$), volumetric domain points ($S_v$), and physical sampling points ($S_s$). The sampled pressure $p(S_s, f)$ constitutes the training data target. \textbf{c}, Architecture of the parallel multi-frequency network and composite loss function. Independent sinusoidal representation network (SIREN) ModMLPs ($\theta_f$) predict the complex pressure $\hat{p}_f$ for each frequency $f$. The total loss $\mathcal{L}$ integrates field fidelity ($\mathcal{L}_\text{data}$: data mismatch at $S_s$; $\mathcal{L}_\text{PDE}$: Helmholtz (partial differential equation) PDE residual at $S_v$) and impedance physics ($\mathcal{L}_\text{var}$: spatial homogeneity of impedance $\zeta$; $\mathcal{L}_\text{smooth}$: frequency smoothness of the spatially-averaged reflection coefficient $\bar{R}$). Impedance $\zeta$ is defined as the normalized surface impedance $\zeta = Z/(\rho_0 c_0)$, where $\rho_0$ and $c_0$ denote the air density and sound speed. It is inferred implicitly via the boundary condition $j\zeta \frac{\partial p}{\partial n} = k p$, with $k=2\pi f/c_0$, using automatic differentiation.}
    \label{fig:framework}
\end{figure*}

The proposed framework for surface impedance inference is based on a parallel neural network architecture, schematically illustrated in \autoref{fig:framework}. The general concept (\autoref{fig:framework}a) involves learning the representation of the steady-state acoustic pressure field patch from sparse microphone measurements taken in the vicinity of the material, where the sound field is typically excited by a single source. The architecture (\autoref{fig:framework}c) employs a set of independent networks, each dedicated to a set of specific frequencies ($f_1, \dots, f_n$) with their own weights ($\theta_1, \dots, \theta_n$). All networks operate on the shared spatial domain defined by three coordinate sets (\autoref{fig:framework}b): boundary coordinates ($S_b$), volumetric domain coordinates ($S_v$), and physical sampling coordinates ($S_s$). Specifically, the physical sampling coordinates correspond to where the steady-state field is sampled, arranged as two parallel layers defined by the surface-to-first-layer distance $d_1$ and the inter-layer spacing $d_2$. These networks are trained simultaneously in a single process, where inference is performed directly during optimization, guided by a shared composite loss function $\mathcal{L}$. This composite loss consists of a data fidelity term ($\mathcal{L}_\text{data}$) minimizing the network's prediction error at sampling positions, a PDE residual loss ($\mathcal{L}_\text{PDE}$) enforcing the source-free Helmholtz equation governing the field, and priors ($\mathcal{L}_\text{var}, \mathcal{L}_\text{smooth}$) that impose spatial homogeneity and frequency smoothness on the inferred impedance.

%-------------------------------%
\subsection{Numerical verification and sensitivity analysis}\label{numerical_verification}
\begin{figure*}[htbp]
    \centering
     \includegraphics[width=\linewidth]{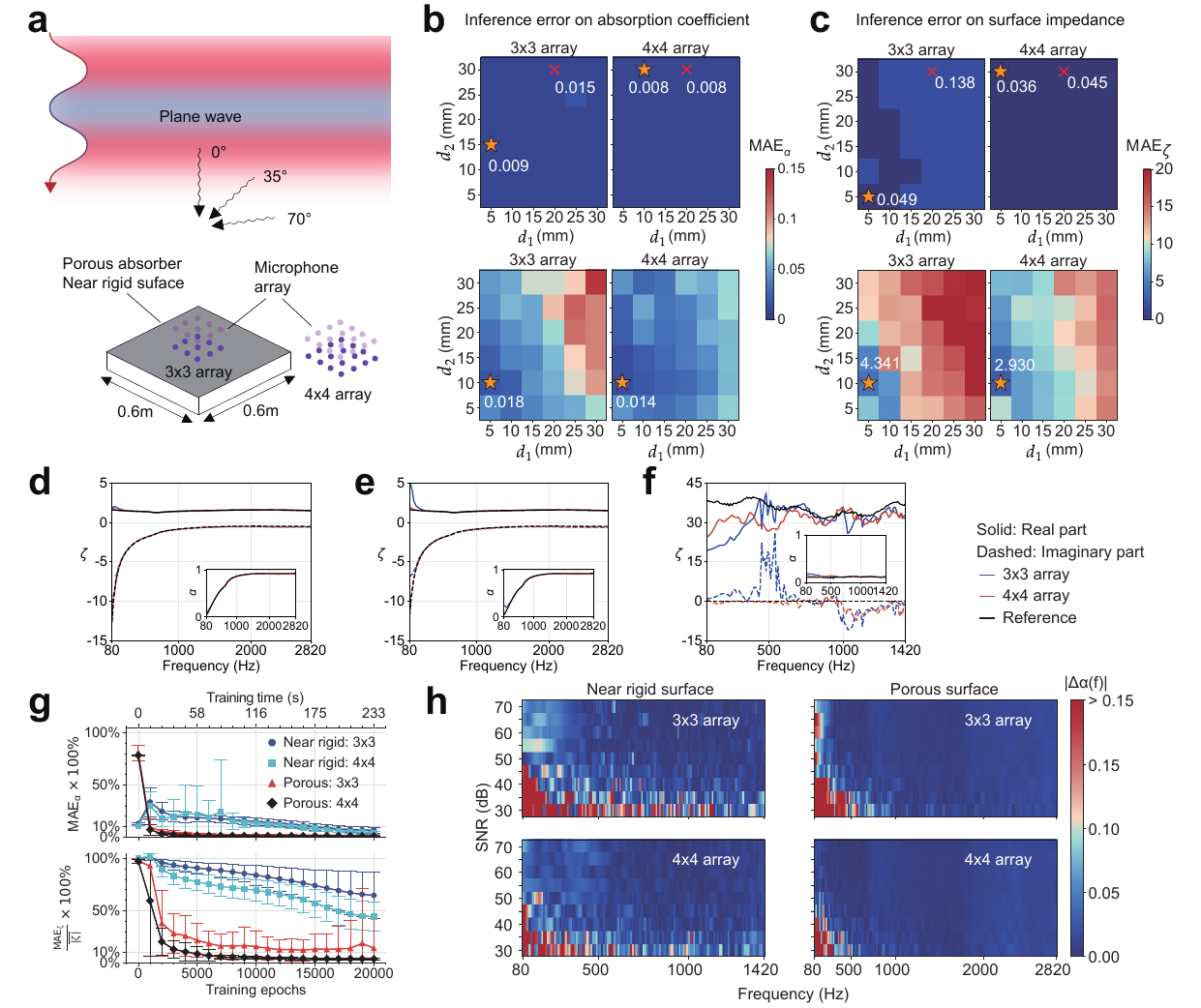}
\caption{\textbf{Numerical verification under plane-wave incidence.} 
\textbf{a}, Simulation setup illustrating plane-wave incidence on a test surface under anechoic conditions, showing the two-layer sparse microphone array geometry.
\textbf{b}, Parametric error analysis for the absorption coefficient $\alpha$. Heatmaps show the \emph{frequency-averaged} mean absolute error ($\mathrm{MAE}_{\alpha}$) as a function of $(d_1,d_2)$ for the porous absorber (top) and the near-rigid surface (bottom), for 3$\times$3 and 4$\times$4 arrays. Stars mark the minimum-error configuration for each array, and the red cross indicates the practically selected configuration used for the porous absorber in subsequent analyses. 
\textbf{c}, Parametric error analysis for the normalized surface impedance $\zeta$. Heatmaps show the \emph{frequency-averaged} mean absolute error ($\mathrm{MAE}_{\zeta}$) under the same geometric sweep and array configurations. 
\textbf{d--f}, Broadband inference results. 
\textbf{d}, Porous absorber using the minimum-impedance-error (starred) configurations. 
\textbf{e}, Porous absorber using the selected practical configuration (red cross). 
\textbf{f}, Near-rigid surface using the minimum-error (starred) configurations. 
\textbf{g}, Training convergence. Evolution of $\mathrm{MAE}_{\alpha}$ (top) and the normalized impedance error $\mathrm{MAE}_{\zeta}/\overline{|\zeta|}$ (bottom) versus training epochs (bottom axis) and wall-clock time (top axis), error bars denote the standard deviation across all investigated $(d_1,d_2)$ configurations. 
\textbf{h}, Sensitivity to additive noise. Heatmaps show the frequency-resolved absolute absorption error $|\Delta \alpha(f)| = |\alpha_{\mathrm{pred}}(f)-\alpha_{\mathrm{gt}}(f)|$ versus frequency and SNR for the selected configurations (porous: red cross, near-rigid: star), with columns corresponding to the 3$\times$3 and 4$\times$4 arrays.}
    \label{fig:plane_wave}
\end{figure*}

We first benchmarked the framework under idealized plane-wave incidence within anechoic conditions (\autoref{fig:plane_wave}a). We performed a parametric sweep across two-layer rectangular microphone arrays with a fixed microphone spacing of 25 mm, 3$\times$3 and 4$\times$4, by varying $d_1$ and $d_2$ for two contrasting boundaries: a porous absorber and a near-rigid surface. Two metrics, $\mathrm{MAE}_{\alpha}$ and $\mathrm{MAE}_{\zeta}$, see \autoref{sec:evaluationmetrics}, were used to quantify the frequency-averaged errors in absorption and impedance, respectively. They are evaluated across frequency ranges of 80–2820 Hz for the porous case and  80–1420 Hz for the near-rigid case.

The parametric analysis (\autoref{fig:plane_wave}b-c) shows that the porous absorber exhibits a largely flat error landscape, indicating high robustness to sensor placement. For the porous material, the 3$\times$3 array shows a slight improvement from closer placement, whereas the 4$\times$4 array retains significantly low error over a wider range, with the lowest error at $d_2 = 30$ mm for both absorption and impedance. Interestingly, the minima for absorption and impedance errors do not coincide. In contrast, the near-rigid material exhibits a highly sensitive error landscape, with optimal performance limited to a closer sensor configuration ($d_1 = 5$ mm, $d_2 = 10$ mm). Moving away from this specific configuration results in rapid accuracy degradation. Such a high sensitivity arises because the normal pressure gradient vanishes near a rigid boundary, significantly reducing the SNR of the physical gradients the network can use. 

In reality, absorbing materials, e.g., automotive seats or trim, are often curved and rather small. Therefore, placing the microphone array at a short distance, such as 5 mm, is not always feasible. We therefore further evaluated the impact of increasing $d_1$ to 20 mm, indicated by the red cross. For the $4{\times}4$ array, this change does not affect the absorption error and increases the impedance error only marginally. However, for the $3{\times}3$ array, the same change leads to a noticeable performance degradation. Given that the $4{\times}4$ array maintains low errors at $d_1$ = $20$ mm, we selected the configuration, $d_1{=}20~\mathrm{mm},\, d_2{=}30~\mathrm{mm}$ for the subsequent analysis. For the near-rigid material, the error landscape offers no flexibility: optimal performance is confined to the near field, with $d_1 = 5$ mm and $d_2 = 10$ mm. Since rigid samples are typically larger and flatter, we adhered to this configuration.

With these configurations, \autoref{fig:plane_wave}d–f compares inferred impedance and absorption to the ground truth. For the porous absorber at the starred impedance configuration (\autoref{fig:plane_wave}d), both arrays closely match the reference across the investigated frequency range. At the red cross configurations (\autoref{fig:plane_wave}e), the 4$\times$4 array maintains excellent agreement, while the 3$\times$3 array deviates particularly in the imaginary impedance below 200 Hz. The difference is largest for the near-rigid case (\autoref{fig:plane_wave}f): the 4$\times$4 array yields somewhat over-estimated real part and a stable near-zero imaginary part, whereas the 3$\times$3 result deviates largely below 600 Hz, with fluctuating imaginary part. These results confirmed that porous absorbers permit sparser spatial sampling, whereas near-rigid boundaries necessitate extended spatial apertures. In all cases, regardless of the absolute impedance error, the residual discrepancy in the absorption coefficient is concentrated at frequencies below 300 Hz, and the error manifests predominantly as an overestimation relative to the reference.

To assess the balance between computational cost and inference accuracy, \autoref{fig:plane_wave}g shows the error as a function of training epochs and time. For the porous absorber, the convergence is rapid and monotonic, reaching a low-error plateau within 2,500–5,000 epochs and 30–60 s across the configurations. The near-rigid case demonstrated a slower convergence, requiring longer training. Notably, in the initial training phase, the absorption error exhibits significant fluctuations, while the impedance error remains consistently high with minimal variance. This indicates that a low absorption error is not a sufficient indicator of accurate impedance estimations, highlighting the non-uniqueness of the inverse mapping from absorption to surface impedance. 

The sensitivity to noise was evaluated in \autoref{fig:plane_wave}h by adding complex Gaussian noise at SNRs ranging from 30 to 70 dB to the noise-free signals. While the absolute error predictably increases with higher noise levels, the degradation profiles diverge significantly between the two boundary types. Specifically, the porous absorber's characterization appears notably less susceptible to noise than that of the near-rigid surface, resulting in lower errors across most SNR conditions. This robustness is attributed to the larger spatial gradients in both magnitude and phase between the two sensor layers for absorptive material. Consequently, with the same absolute level of additive noise, the relative contamination of the true pressure difference is smaller. As expected, the 4$\times$4 array is more tolerant to noise, particularly at SNRs below 50 dB. Notably, for SNRs below 40 dB, the smaller array exhibits pronounced low-frequency errors for the near-rigid material, whereas the 4$\times$4 array maintains a stable plateau.

%-------------------------------%
\subsection{Experimental validation on finite-sized specimens}\label{expeimental_validation}
\begin{figure*}[htbp]
    \centering
 \includegraphics[width=\linewidth]{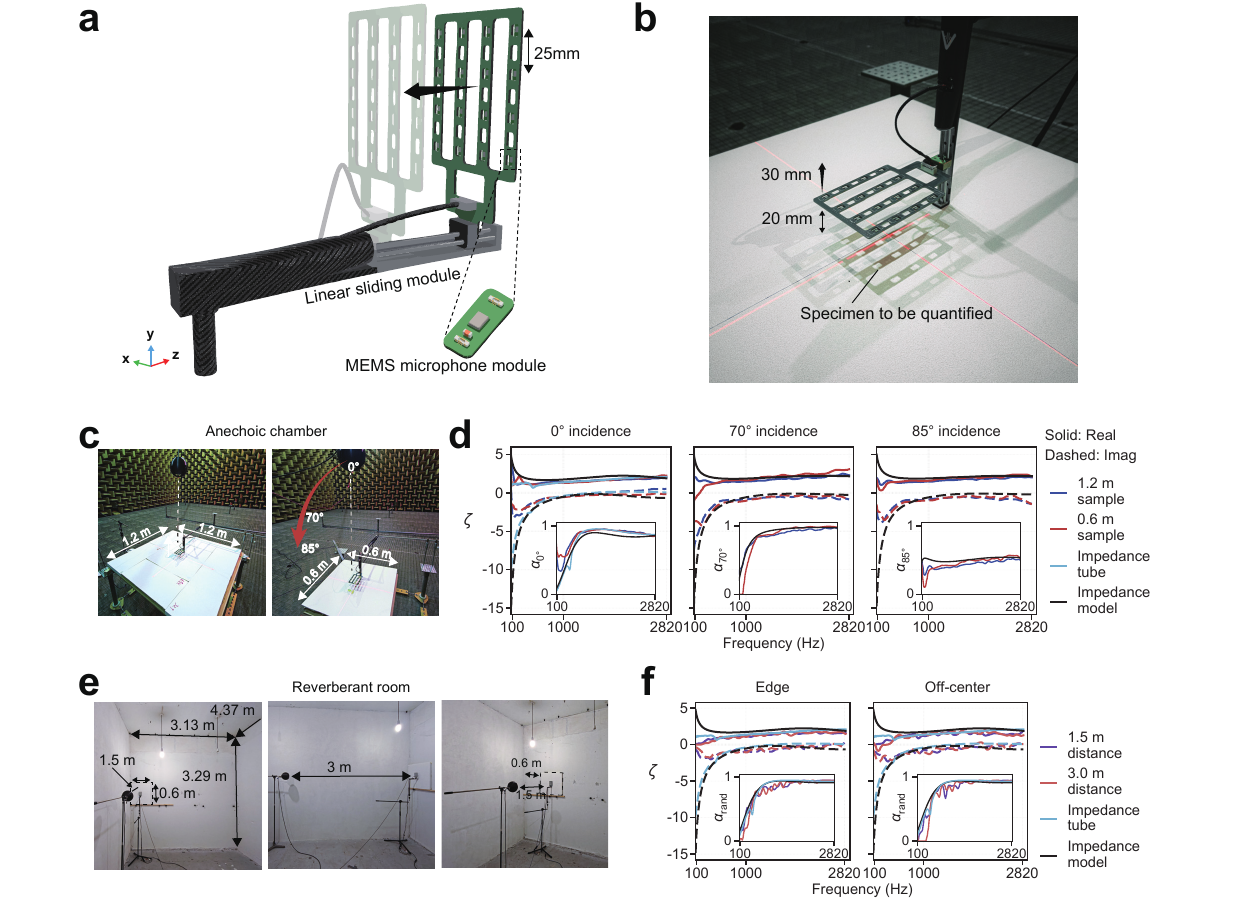}
\caption{\textbf{Experimental validation on finite-sized specimens across reflection-free and reverberant environments.} \textbf{a}, Illustration of the custom-built microphone array featuring a linear sliding module and a $4\times4$ Micro-electro-mechanical systems (MEMS) microphone grid. \textbf{b}, Photograph of the experimental deployment showing the array positioned above a specimen with defined sampling heights ($d_1=20$ mm, $d_2=30$ mm). \textbf{c}, Anechoic chamber setup for characterizing angle-dependent properties. Measurements were conducted on square specimens of two edge lengths ($0.6$ m and $1.2$ m) under normal ($0^\circ$), oblique ($70^\circ$), and grazing ($85^\circ$) incidence. \textbf{d}, Inferred surface impedance $\zeta$ (solid/dashed lines for real/imaginary parts) and absorption coefficients $\alpha$ (insets) for the anechoic cases, compared against impedance tube data and theoretical models. \textbf{e}, Reverberant chamber setup testing the specimen at edge and off-center positions with varying source distances (1.5 m and 3.0 m) to modulate the direct-to-reverberant ratio. \textbf{f}, Inferred results in the reverberant environment: inferred impedance and random incidence absorption coefficients compared with reference values from impedance tube (measured normal incidence impedance and random incidence absorption coefficient derived via Paris' formula~\cite{paris1928coefficient}) and impedance model.}
\label{fig:exp_validation}
\end{figure*}

We developed a custom-built microphone array featuring a 4$\times$4 grid with removable microphone modules and a linear sliding module (\autoref{fig:exp_validation}a) to perform experimental validation. We first conducted experiments in an anechoic chamber to validate the inference of angle-dependent impedance using the setup shown in \autoref{fig:exp_validation}c. We utilized a square porous absorber, Ecophon Master A, with a thickness of 40 mm, as the test specimen. Measurements cover incidence angles of $\theta_{\text{inc}} = 0^\circ, 70^\circ, 85^\circ$ for square specimens with widths of 0.6 m and 1.2 m. The angle-dependent references are modeled with Qunli parameters~\cite{qunli1988empirical} with a flow resistivity value $\sigma = 36,790~\mathrm{Pa \cdot s \cdot m^{-2}}$ provided by the manufacturer. The normal incidence reference data were also provided by the manufacturer, based on their impedance tube measurements. We used the configuration with $d_1=20$ mm and $d_2=30$ mm and set the training duration to 2,500 epochs for these scenarios.

As shown in \autoref{fig:exp_validation}d, compared to the reference curve, the small specimen with a side length of 0.6 m exhibits distinct absorption coefficient bumps that shift towards higher frequencies relative to the 1.2 m specimen. This phenomenon corresponds to edge diffraction effects, which occur when the acoustic wavelength becomes comparable to the object's size and is visible in most anechoic chamber measurements, see, e.g.,~\cite{ottink2016situ,richard2017estimation}. Although the array only captures the local pressure field at the center of the specimen, the framework successfully reconstructs the local sound field, enabling the detection of the non-planar wavefront curvature induced by edge effects. 

Interestingly, the edge diffraction effect is reduced as the incident angle increases. At $70^\circ$ incidence angle, the inferred impedance for the 1.2 m specimen nearly matches the impedance model. At a near-grazing incidence of $85^\circ$, these diffraction phenomena further diminish, as the incident wave propagates nearly parallel to the surface, reducing the back-scattering of energy from the far edges to the array center. In this regime, the framework accurately retrieves the real part, while the deviation in the imaginary part below 200 Hz is attributed to the vanishing normal particle velocity. At this limit, the minute phase gradient between the two layers becomes comparable to the MEMS microphones' inherent phase noise floor, leading to inaccurate estimation. For both specimen sizes, minor deviations in the inferred impedance and absorption coefficient were observed at higher frequencies under oblique incidence compared to the impedance model. These discrepancies are likely attributable to the speaker alignment tolerances and inaccuracies in the incident angle.

Following free-field validation, the framework was evaluated in a reverberant chamber (\autoref{fig:exp_validation}e), restricted to the 0.6 m specimen. The chamber is characterized by strong, multi-directional reflections with a broadband reverberation time ($T_{20}$) of 6.3 s. To evaluate the influence of material placement, we tested two distinct placement configurations: an ``edge'' position, where the specimen's edge meets the room's edge, and an ``off-center'' position, located between the room center and the edge. Compared to the anechoic condition results, the inferred impedance in the reverberant setting exhibits increased fine-scale fluctuations, mainly due to the complex reflection patterns. Nevertheless, despite these deviations, the inferred results show strong agreement with the spectral trends of the reference, irrespective of the source distance. This consistency extends to the 3.0 m case, where the direct field is significantly out-powered by the reverberant energy, the data remain consistent with the results at 1.5 m (\autoref{fig:exp_validation}f). Such consistent results are obtained because the neural network focuses on the local sound field near the wall and is therefore insensitive to the global sound field properties.

Furthermore, the edge configuration yields notably smoother spectral inference than the off-center position. Such a superior performance is attributed to the corner boundaries acting as acoustic mirrors, which increase the apparent size of the tested specimen and suppress the edge diffraction artifacts observed in free-field conditions. Additionally, the edge serves as a pressure accumulation zone, naturally maximizing the SNR for the measurement~\cite{waterhouse1958output,berzborn2025directional}. In contrast, the off-center results exhibit low-frequency fluctuations driven by the interference of strong early reflections, which create a more complex local modal field. Despite these environmental complexities and the previously noted inherent phase noise affecting the imaginary part, the inference of the real part remains robust relative to the reference, and the random incidence absorption coefficients are accurately estimated. This indicates that when the SNR is lower, the accuracy of the inferred imaginary part is more susceptible to noise; however, it converges to a value that, combined with the accurate real part, still yields a reasonable absorption coefficient.

%-------------------------------%
\subsection{Local field complexity indicates inference accuracy}\label{local_field}
\begin{figure*}[htbp]
    \centering
    \includegraphics[width=\linewidth]{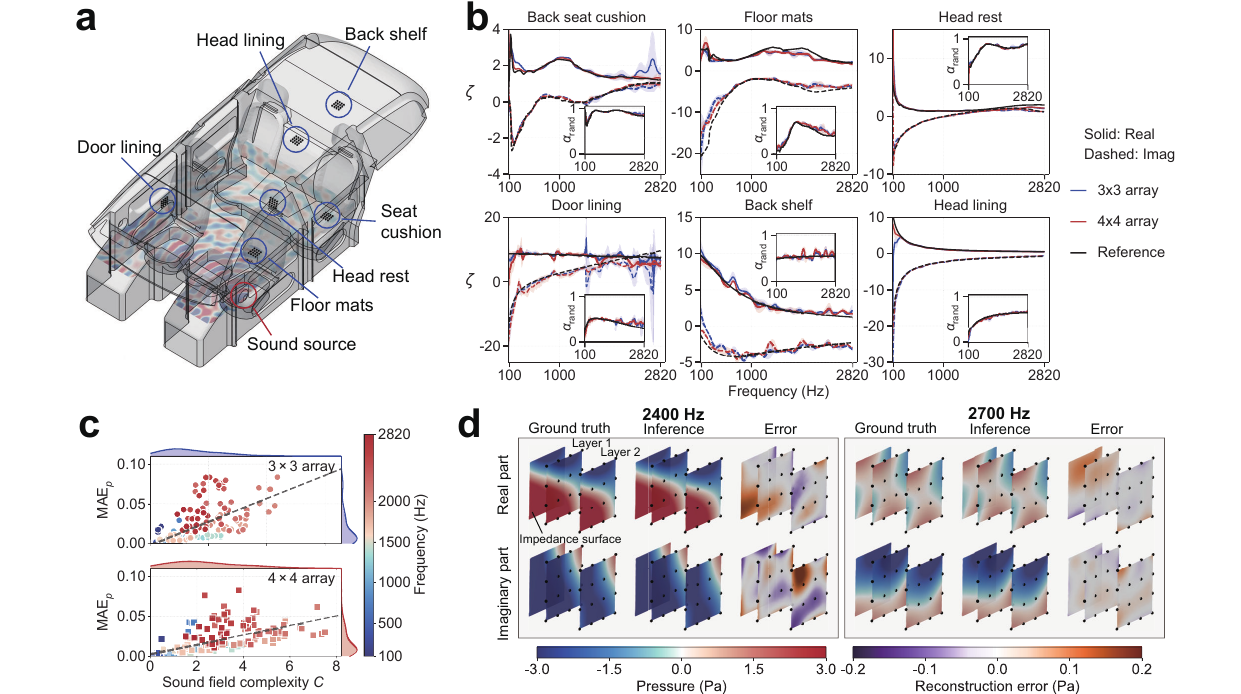}
\caption{\textbf{Virtual \emph{in-situ} characterization of multi-component acoustics in a full vehicle model.} \textbf{a}, 3D visualization of the vehicle cabin sound field at 2800 Hz, highlighting the target interior components for impedance characterization. \textbf{b}, Comparison of the inferred broadband impedance versus ground truth references for various interiors using $3\times3$ and $4\times4$ array configurations. Shaded regions represent the standard deviation across multiple spatially distributed test positions ($N=2\text{--}4$) for each component, indicating spatial consistency. Insets display the corresponding absorption coefficients. \textbf{c}, Correlation between the prediction error of the pressure ($\mathrm{MAE}_p$) and the sound field complexity metric ($C$) at the door lining location (evaluated within a 20 mm height volume consisting of 52,500 points above the surface). Data are shown for $3\times3$ (circles) and $4\times4$ (squares) arrays, color-coded by frequency. \textbf{d}, Visualization of ground truth pressure field slices, network predictions, and corresponding error maps (real and imaginary parts) at 2400 Hz and 2700 Hz near the door lining, illustrating the two sampling layers and the impedance surface. The error maps show that sharp nodal line structures result in high-complexity regions, leading to increased inference error at the target impedance surface.}
\label{fig:vehicle}
\end{figure*}

While experimental validation confirms the framework's practical feasibility, quantifying and analyzing the factors that influence its accuracy in complex environments remains challenging due to the lack of ground truth data. To address this challenge and demonstrate the network's applicability in a practical use case, we conducted a virtual \emph{in-situ} characterization within the acoustic cavity of a full vehicle model (\autoref{fig:vehicle}a). The sound field was generated by a virtual speaker source mounted on the door trim, configured to mimic realistic automotive conditions with an SNR of 70 dB. This ensures that extrapolation errors are primarily attributable to local sound field complexity rather than measurement noise.

We performed impedance characterization for multiple key interior components, ranging from absorptive elements, such as seat cushion, floor mats, and headlining, to acoustically reflective surfaces in complex interference zones, e.g., headrest, door lining, and back shelf (\autoref{fig:vehicle}b). Although we anticipate degraded performance with the smaller array, we employed both array sizes for these investigations. This choice is motivated by the difficulties of measuring in confined cabin spaces with a larger array, while also helping to better understand the underlying mechanisms of performance degradation. In this section, except for the door lining component, an adaptive training strategy that determines the termination of training epochs based on the pressure gradient was implemented.

To test spatial consistency, simulated measurements were repeated at multiple spatial offsets for each component, by shifting the array along one axis across 2–4 discrete locations while holding the other fixed. In \autoref{fig:vehicle}b, the shaded regions denote the standard deviation, while the solid dashed curves represent the mean values. For generally absorptive components, the inferred broadband impedance shows good agreement with the ground truth for both 3$\times$3 and 4$\times$4 array configurations, indicating high consistency across sampling locations. However, for reflective surfaces, such as the back shelf and door lining, predictions exhibit larger deviations with varying locations, particularly above 2 kHz. For the door lining, which is identified as a high-complexity zone, a finer sampling configuration, $d_1=5$ mm, $d_2=10$ mm, and an extended training up to 20,000 epochs were applied. Despite these enhanced settings, deviations persist, signifying intrinsic physical challenges in these regions.

To understand the physical causes of the challenges, we analyzed the relationship between the prediction error of the pressure ($\mathrm{MAE}_p$) and the local sound field complexity metric $C$ (\autoref{eq:complexity}) computed over a volumetric point set $\mathcal{E}$ spanning 20 mm above the impedance surface, specifically for the door lining (\autoref{fig:vehicle}c). A clear correlation is observed. The higher the sound field complexity, the higher the pressure prediction errors. The scatter plot also shows that, while the $4\times4$ array achieves a lower error baseline than the $3\times3$ configuration, the upward trend with complexity remains the same. Notably, the color maps indicate that the highest sound field complexity values, and consequently the largest errors, cluster predominantly at frequencies above 2 kHz, suggesting that the challenge stems from increasingly intricate field complexity at shorter wavelengths.

The specific complexity features driving these errors are visualized in \autoref{fig:vehicle}d. Here, we find that extrapolation fidelity is limited not only by frequency but also by the presence of strong nodal lines. By comparing sound fields at two frequencies with similar wavelengths, 2400 Hz and 2700 Hz, a distinct contrast emerges. At 2400 Hz, the ground truth reveals a sharp nodal line structure, characterized by vanishing pressure amplitude and rapid phase transitions. This sound field complexity challenges the sparse array's capabilities, resulting in significant extrapolation errors distributed across the entire sampling patch, where strong patterns in the real and imaginary parts are coupled and mutually influential. In contrast, the sound field at 2700 Hz exhibits a smoother distribution, without such sharp gradients, which the network can reconstruct with higher fidelity using the same sensor configuration. This comparison confirms that the prediction bottleneck can be the local field structure itself: the presence of nodal lines increases the local patch's global complexity, thereby degrading overall extrapolation accuracy.

%-------------------------------%
\section{Discussion}\label{discussion}
Our numerical and experimental analyses reveal that the performance of the proposed material characterization framework is mainly affected by two factors: the acoustic properties of the target material and the local sound field complexity near the surface. As demonstrated in the numerical verification, the conditioning of the inverse problem is intrinsically linked to the material nature. For absorbing materials, significant energy dissipation induces strong pressure gradients across the sensor layers, ensuring sufficient SNR and providing informative gradients for the physics-informed loss. This explains the rapid convergence within 5,000 epochs and high noise tolerance observed in simulations, as well as the accurate characterization of absorptive interiors in the virtual vehicle cabin. Conversely, near-rigid surfaces and high-angle oblique incidence present a greater challenge: the vanishing particle velocity results in weak pressure gradients, rendering the inversion process highly sensitive to measurement noise. This necessitates significantly longer training, up to approximate  $20,000$ epochs, to resolve the subtle pressure differences, yet extrapolating these weak gradients to the impedance surface remains an ill-posed task, ultimately limiting inference accuracy.

Despite the framework's ability to operate with sparse data, the observed material-dependent performance highlights the critical role of a larger sampling area. Our parametric sweeps show that increasing the array size from 3$\times$3 to 4$\times$4 acts as a powerful spatial regularizer. By providing additional measurement data to constrain the solution within the volumetric domain, the larger array imposes stronger constraints on the solution space, suppressing fluctuations on reflective surfaces and mitigating errors induced by increased noise at lower frequencies. This spatial regularization is equally vital in experimental scenarios, as it facilitates the capture of wavefront curvature induced by edge diffraction, while in the vehicle cabin, it enhances robustness against complex interference patterns.

Furthermore, the results clarify the impact of the acoustic environment. While anechoic environments are conventionally preferred to avoid reflections from the boundaries of the room, we found that reverberant environments can actually facilitate measurement by creating a more distributed sound incidence, although the field is not strictly diffuse, which helps reduce the impact of the diffraction effect, particularly diffraction from the edges of the finite sample, for evaluating the absorption coefficient. The primary limitation in these enclosed spaces arises not from global reverberation but from local field complexity, especially near nodal lines where the pressure vanishes and phase changes rapidly. 

As identified in the vehicle study, these regions of high complexity coincide with peak prediction errors. Building on this insight, we propose a practical guideline for \emph{in-situ} measurements in enclosed semi-reverberant spaces, e.g., vehicle cabins or small rooms, whenever structurally feasible and maintaining a sufficiently low background noise, the acoustic enclosure should preferably be ``opened'' to the external environment, such as by opening doors, windows, or trunk. This intervention disrupts the formation of strong standing waves. By physically preventing the buildup of stable resonant patterns, the occurrence of sharp nodal lines is significantly reduced. Consequently, the local sound field becomes smoother, effectively regularizing the inverse problem and preventing the underlying mathematical formulation from becoming ill-posed in regions of vanishing sound pressure. This maximizes the extrapolation accuracy of the physics-informed framework without requiring denser microphone arrays.

In summary, the proposed framework offers a reliable means for rapid, \emph{in-situ} acoustic material characterization. Looking ahead, future work aims to integrate active sampling strategies that quantify complexity zones and frequency regions to better adjust microphone placement. Beyond these algorithmic refinements, the framework is designed to scale with ongoing technological advancements. As MEMS sensors evolve toward miniaturization and higher SNR, the precision of input data will naturally improve, mitigating current limitations such as high inherent noise at low frequencies. Simultaneously, the ongoing advances in GPU acceleration and deep learning infrastructure will further reduce computational costs and latency. Consequently, the proposed methodology represents a forward-looking paradigm that grows in efficacy alongside these parallel advancements in hardware and computing power.

%-------------------------------%
\section{Methods}\label{methods}
\subsection{Neural field formulation and composite loss design}
The parallel neural field architecture consists of a number of frequency channels in a SIREN-based modified multilayer perceptron (ModMLP)~\cite{sitzmann2020siren,wang2021understanding}, parameterized independently for each frequency and evaluated in parallel across frequencies. It can be equivalently viewed as $\{ \Phi_i(\mathbf{x}; \theta_i) \}_{i=1}^{N}$, where $N$ is the number of frequency bins. Each channel $\Phi_i$ maps a 3D spatial coordinate $\mathbf{x} \in \mathbb{R}^3$ to the predicted real and imaginary parts of the complex-valued pressure $\hat{p}_i(\mathbf{x})$ at the $i$-th discrete frequency. All channels share the same spatial sampling sets for the boundary coordinates ($S_b$), volumetric domain coordinates ($S_v$), and sampling coordinates ($S_s$), as defined in \autoref{fig:framework}b.

The training objective is governed by a composite, physics-informed loss function $\mathcal{L}(\theta)$, where $\theta = \{\theta_i\}_{i=1}^{N}$ represents the trainable parameters. This loss aggregates weighted contributions from data, physics, and regularization constraints. We formulate the total loss as:
\begin{equation}
\mathcal{L}(\theta)=
\sum_{i=1}^{N}\Big(
\lambda^{\text{data}}_{i}\,\mathcal{L}^{\text{data}}_{i}
+
\lambda^{\text{PDE}}_{i}\,\mathcal{L}^{\text{PDE}}_{i}
+
\lambda^{\text{var}}_{\mathrm{r},i}\,\mathcal{L}^{\text{var}}_{\mathrm{r},i}
+
\lambda^{\text{var}}_{\mathrm{i},i}\,\mathcal{L}^{\text{var}}_{\mathrm{i},i}
\Big)
+
\lambda^{\text{smooth}}\,\mathcal{L}^{\text{smooth}},
\end{equation}
\noindent where the variance terms correspond to the real and imaginary parts of the inferred normalized impedance $\zeta_i(\mathbf{x})$ at the boundary, and $\lambda^{(\cdot)}$ are weighting coefficients. The individual loss components are defined as follows:
\begin{enumerate}
    \item Data target: Minimizes the discrepancy between the predicted pressure $\hat{p}_i(\mathbf{x})$ and the sampling pressure $p_{s,i}(\mathbf{x})$ at sampling coordinates $S_s$:
    \begin{equation}
    \mathcal{L}^\text{data}_{i} =
    \frac{1}{|S_s|}
    \sum_{\mathbf{x} \in S_s}
    \left(
    \left|\mathrm{Re}\!\left\{\hat{p}_i(\mathbf{x})-p_{s,i}(\mathbf{x})\right\}\right|^2
    +
    \left|\mathrm{Im}\!\left\{\hat{p}_i(\mathbf{x})-p_{s,i}(\mathbf{x})\right\}\right|^2
    \right).
    \end{equation}

    \item PDE residual: Enforces the Helmholtz equation $(\nabla^2 + k_i^2)p_i = 0$ for the predicted field $\hat{p}_i$, where $k_i = 2\pi f_i/c_0$ is the wavenumber:
    \begin{equation}
    \mathcal{L}^\text{PDE}_{i} =
    \frac{1}{|S_v \cup S_s|}
    \sum_{\mathbf{x} \in S_v \cup S_s}
    \left(
    \left|\mathrm{Re}\!\left\{\nabla^2 \hat{p}_i(\mathbf{x})+k_i^2 \hat{p}_i(\mathbf{x})\right\}\right|^2
    +
    \left|\mathrm{Im}\!\left\{\nabla^2 \hat{p}_i(\mathbf{x})+k_i^2 \hat{p}_i(\mathbf{x})\right\}\right|^2
    \right)
    \end{equation}
    \noindent evaluated on $S_v \cup S_s$ to regularize the field at measurement locations.

    \item Physical constraints on boundary impedance: Imposes constraints related to the inferred normalized impedance $\zeta_i(\mathbf{x})$ on the boundary $S_b$. The impedance is inferred from the predicted pressure via the linearized Euler equation.
    \begin{equation}
    j\,\zeta_i(\mathbf{x})\,\partial_n \hat{p}_i(\mathbf{x}) = k_i\,\hat{p}_i(\mathbf{x}), \qquad \mathbf{x}\in S_b,
    \end{equation}
    \noindent where $j=\sqrt{-1}$ and $\partial_n(\cdot)\triangleq \mathbf{n}(\mathbf{x})^\top\nabla(\cdot)$ denotes the outward normal derivative with $\mathbf{n}(\mathbf{x})$ pointing outward from the acoustic domain (i.e., out of the sound field region). Accordingly,
    \begin{equation}
    \zeta_i(\mathbf{x}) \triangleq \frac{k_i\,\hat{p}_i(\mathbf{x})}{j\,\partial_n \hat{p}_i(\mathbf{x})},\qquad \mathbf{x}\in S_b.
    \end{equation}
    \begin{itemize}
        \item \emph{Variance loss:} Based on the assumption that the material is locally homogeneous, this term penalizes the spatial variance of the inferred impedance across $S_b$:
        \begin{equation}
        \mathcal{L}^\text{var}_{\mathrm{r},i}
        = \mathrm{Var}_{\mathbf{x}\in S_b}\!\left(\mathrm{Re}\{\zeta_i(\mathbf{x})\}\right),\qquad
        \mathcal{L}^\text{var}_{\mathrm{i},i}
        = \mathrm{Var}_{\mathbf{x}\in S_b}\!\left(\mathrm{Im}\{\zeta_i(\mathbf{x})\}\right).
        \end{equation}

        \item \emph{Reflection coefficient smoothness loss:} Based on the physical prior that acoustic response varies smoothly across frequencies, this global term penalizes the curvature of the spatially-averaged reflection coefficient spectrum.
        First, the spatially-averaged impedance $\bar{\zeta}_i$ is computed for each frequency $i$:
        \begin{equation}
        \bar{\zeta}_i = \frac{1}{|S_b|} \sum_{\mathbf{x} \in S_b} \zeta_i(\mathbf{x}).
        \end{equation}
        The corresponding spatially-averaged reflection coefficient is $\bar{R}_i = \frac{\bar{\zeta}_i-1}{\bar{\zeta}_i+1}.$
        
        The loss minimizes the second-order finite difference $\Delta^2 \bar{R}_{i} = \bar{R}_{i+1} - 2\bar{R}_{i} + \bar{R}_{i-1}$ using the Huber loss~\cite{huber1964robust} ($\mathcal{L}_{\delta}$ with $\delta=0.5$):
        \begin{equation}
        \mathcal{L}^\text{smooth} =
        \frac{1}{N-2}
        \sum_{i=2}^{N-1}
        \left[
        \mathcal{L}_{\delta}\!\left(\mathrm{Re}(\Delta^2 \bar{R}_{i})\right)
        +
        \mathcal{L}_{\delta}\!\left(\mathrm{Im}(\Delta^2 \bar{R}_{i})\right)
        \right].
        \end{equation}
    \end{itemize}
\end{enumerate}

Gradient computation for all terms involving spatial derivatives is performed via automatic differentiation using PyTorch 2.9.0~\cite{paszke2019pytorch}. In particular, spatial derivatives (Laplacian and boundary normal derivatives) are obtained by forward-mode JVP, and the frequency-parallel evaluation uses tensor contraction to apply per-frequency weights efficiently.

\subsection{Training protocol}\label{training_protocol}
The architecture reduces the computational burden of broadband wavefield modeling by training one network per frequency bin in parallel, such that each network independently represents a local continuous sound field patch. Prior to training, we perform frequency-wise preprocessing to stabilize optimization: for each frequency bin, complex pressures across sensors are normalized by the maximum magnitude, phase-aligned to a reference sensor chosen as the one with the largest magnitude at that frequency, and split into real and imaginary components.

During training, we employ a self-adaptive weighting strategy to dynamically balance different loss terms. Specifically, we update the weights $\lambda_{q,i}$ for each loss component $\mathcal{L}_{q,i}$, where $q\in\{\text{data},\text{PDE},\text{var},\text{smooth}\}$ denotes the loss type and $i$ denotes the frequency bin, at each weight-update step $t$, performed every 100 epochs of optimizer iterations. Target weights $\hat{\lambda}_{q,i}^{(t+1)}$ are computed from the $\ell^2$-norm of the per-frequency gradient magnitudes
\begin{equation}
G_{q,i}^{(t)}=\left\|\nabla_{\theta_i}\mathcal{L}_{q,i}\right\|_2,
\end{equation}
relative to the combined data-and-residual scale.

For the data and PDE residual losses, we define $S_i^{(t)}=G_{\text{data},i}^{(t)}+G_{\text{PDE},i}^{(t)}$ and set
\begin{equation}
\hat{\lambda}_{\text{data}, i}^{(t+1)}=\frac{S_i^{(t)}}{G_{\text{data}, i}^{(t)}},
\qquad
\hat{\lambda}_{\text{PDE}, i}^{(t+1)}=\frac{S_i^{(t)}}{G_{\text{PDE}, i}^{(t)}}.
\end{equation}
For regularization terms, the target scale $S_i^{(t)}$ includes the corresponding gradient magnitudes, ensuring that the regularization strength is commensurate with the physics and data constraints.

Finally, the weights are updated using an exponential moving average:
\begin{equation}
\lambda_{q,i}^{(t+1)}=\alpha_q \lambda_{q,i}^{(t)} + (1-\alpha_q)\hat{\lambda}_{q,i}^{(t+1)},
\end{equation}
where $\alpha_q \in (0,1)$ are loss-specific smoothing factors for $q\in\{\text{data},\text{PDE},\text{var},\text{smooth}\}$. We set $\alpha_{\text{data}}=\alpha_{\text{PDE}}=0.90$ and $\alpha_{\text{var}}=0.999$. For the smoothness term, we set $\alpha_{\text{smooth}}=1-\frac{10000}{E^2}$, where $E$ is the total epoch budget.

Specifically for the local field reconstruction analysis (\autoref{local_field}), we implement an equation to dynamically determine the total number of training epochs. An input complexity index, denoted as $I$, was computed based on the magnitude of the spatially averaged complex pressure difference between the two sampling planes:
\begin{equation}
I = \left\langle \frac{|\langle \Delta p \rangle_{\mathbf{x}}|}{k \cdot d_2 \cdot \langle |p| \rangle_{\mathbf{x}}} \right\rangle_f \times 10^{4},
\end{equation}
where $\langle \cdot \rangle_{\mathbf{x}}$ denotes spatial averaging across the sensor plane and $\langle \cdot \rangle_f$ denotes averaging over discrete frequency. The term $\langle |p| \rangle_{\mathbf{x}}$ represents the spatial mean of the pressure magnitude. The resulting index $I$ was then mapped to a discrete epoch budget selected from the set \{1000, 2500, 5000, 10000\} using empirically defined thresholds.

We adopt the SOAP optimizer for all experiments~\cite{vyas2024soap}, which preconditions updates via gradient covariance with preconditioning performed every two optimizer steps to enhance stability and convergence in the ill-posed inverse problem. A warmup-cosine learning rate schedule was employed. All training presented in this paper was conducted on a single Nvidia RTX 5090.

\subsection{Numerical simulation setups}\label{numerical_set}
Full-wave acoustic simulations were performed using the Pressure Acoustics Frequency Domain module within the finite element software COMSOL Multiphysics\textsuperscript{\textregistered} (v6.3)~\cite{COMSOL_v6_3}. Standard air properties were assumed for all simulations (air density $\rho_0 = 1.2~\mathrm{kg\cdot m^{-3}}$, speed of sound $c_0 = 343.2~\mathrm{m\cdot s^{-1}}$). Perfectly matched layers were employed to simulate free-field conditions in the plane-wave incidence studies (\autoref{fig:plane_wave}). The mesh resolution was set based on the highest frequency of interest. For the anechoic environment simulations, the maximum element size was 0.0142 m (approx. 9 points per wavelength (PPW) at 2820 Hz), while for the vehicle cabin simulation (\autoref{fig:vehicle}), it was 0.0095 m (approx. 13 PPW at 2820 Hz). Plane wave excitation was implemented using a background pressure field. For the vehicle cabin simulation, sound sources were modeled as normal-velocity boundary conditions applied to specific interior surfaces, configured so that the source strength corresponds to a free-field pressure of 2.5 Pa at 1 m.

Two material types were simulated under plane wave incidence. The near-rigid surface (20 mm thick) was modeled using an impedance boundary condition assuming locally reacting behavior with a nearly constant normalized impedance $\zeta$ where $\mathrm{Re}(\zeta) \approx 38-40$ and $\mathrm{Im}(\zeta) = 0$. The porous absorber (40 mm thick, details in \autoref{fig:plane_wave} caption) was simulated using the Poroacoustics module with the Miki model~\cite{miki1990acoustical}, employing a flow resistivity of $39,260~\mathrm{Pa \cdot s\cdot  m^{-2}}
$. For the vehicle cabin simulation, the geometry model is adopted from the COMSOL Application Gallery~\cite{COMSOL_Car_Cabin_Model}; all interior surfaces were modeled using locally reacting impedance boundary conditions. The impedance data for most interiors were adopted from Aretz et al.~\cite{aretz2014combined}, while data for the back shelf and headlining were approximated based on typical porous material characteristics due to perceived unreliability in the original source. For simulations using the impedance boundary condition, the reference impedance was the input value itself. For the poroacoustics simulation, the reference impedance was determined by calculating the spatially averaged ratio of pressure to normal velocity from the COMSOL results across the boundary coordinates. As the neural network also predicts the spatially averaged impedance, this ensures a fair comparison.

\subsection{Experimental measurement setups}\label{experiment_set}
An in-house designed MEMS microphone array was used for the experiments. Each microphone node incorporates a TDK ICS-43434 microphone (65 dBA SNR) sampling at 24 kHz with 16-bit resolution. Magnitude calibration of the array microphones was performed using an \emph{in-situ} substitution method against a quarter-inch reference microphone (Brüel \& Kjær 4191), whose free-field sensitivity is traceable to the Danish Primary Laboratory of Acoustics (DPLA). Phase calibration was performed relative to the first channel of the array, assuming ideal channel synchronization. A spherical sound source driven by a Scan Speak 10F unit was employed. For measurements intended to approximate plane-wave conditions (e.g., calibration or anechoic chamber), the source was placed 1.2 m from the array.

Measurements were conducted using an exponential sinesweep signal with a duration of 5 s, covering the frequency range from 80 Hz to 4050 Hz. For each sampling position, the measurement was repeated 8 times, and the resulting Fast Fourier Transform (FFT) spectra were averaged in the frequency domain to improve the SNR. The input data for the neural network consisted of approximately 200 linearly spaced frequency bins selected from the averaged spectra between 80 Hz and 2820 Hz. Ambient temperature and humidity were recorded during measurements to allow for accurate calculation of the speed of sound.

\subsection{Evaluation metrics}\label{sec:evaluationmetrics}

Generally, for frequency index $i$, the absorption coefficient at an incidence angle $\theta_{\text{inc}}$ is derived from the spatially averaged normalized impedance $\bar{\zeta}_i$ as:
\begin{equation}
\alpha_{\theta_{\text{inc}},i}=1-\left|\frac{\bar{\zeta}_i\cos(\theta_{\text{inc}})-1}{\bar{\zeta}_i\cos(\theta_{\text{inc}})+1}\right|^2.
\label{eq:alpha}
\end{equation}

For numerical verification, under normal incidence ($\theta_{\text{inc}}=0$), we denote $\alpha_i \triangleq \alpha_{\theta_{\text{inc}}=0,i}$, the mean absolute error is computed as:
\begin{equation}
\mathrm{MAE}_{\alpha}=\frac{1}{N}\sum_{i=1}^{N}\left|\alpha_{i}-\alpha_{\text{ref},i}\right|.
\label{eq:mae_alpha}
\end{equation}

Similarly, the impedance error is quantified by the component-wise mean absolute difference:
\begin{equation}
\mathrm{MAE}_{\zeta}=\frac{1}{2N}\sum_{i=1}^{N}
\left(
\left|\operatorname{Re}\left(\bar{\zeta}_i-\bar{\zeta}_{\text{ref},i}\right)\right|
+
\left|\operatorname{Im}\left(\bar{\zeta}_i-\bar{\zeta}_{\text{ref},i}\right)\right|
\right).
\label{eq:mae_zeta}
\end{equation}

For experimental validation, the random incidence absorption coefficient is estimated from $\bar{\zeta}_i=r_i+jx_i$ using the classical formula for locally reacting surfaces~\cite{paris1928coefficient}:
\begin{equation}
\alpha_{\text{random},i}=\frac{8r_i}{r_i^2+x_i^2}\Bigg[1-\frac{r_i}{r_i^2+x_i^2}\ln\big((r_i+1)^2+x_i^2\big)
+\frac{r_i^2-x_i^2}{x_i(r_i^2+x_i^2)}\arctan\!\left(\frac{x_i}{r_i+1}\right)\Bigg].
\label{eq:alpha_random}
\end{equation}

To quantify the prediction fidelity, let $\mathcal{E}=\{\mathbf{x}_v\}_{v=1}^{|\mathcal{E}|}$ denote the evaluation set. The pressure prediction error at each frequency index $i$ over $\mathcal{E}$ is defined  with respect to the reference pressure $p_{\text{ref},i}(\mathbf{x})$ as:
\begin{equation}
\mathrm{MAE}_{p,i}=\frac{1}{|\mathcal{E}|} \sum_{\mathbf{x} \in \mathcal{E}}
\left(
\left|\operatorname{Re}\left(\hat{p}_i(\mathbf{x})-p_{\text{ref}, i}(\mathbf{x})\right)\right|
+
\left|\operatorname{Im}\left(\hat{p}_i(\mathbf{x})-p_{\text{ref}, i}(\mathbf{x})\right)\right|
\right).
\label{eq:mae_p}
\end{equation}

To characterize the local sound field variation, a sound field complexity metric is introduced based on the spatial variation of the pressure field. Specifically, we define this metric using the mean absolute first difference of the pressure values evaluated on the set $\mathcal{E}$. The points in  $\mathcal{E}$ are generated via Latin-hypercube sampling within the local volumetric domain above the impedance surface, and are ordered sequentially based on their generation index to form a sequence 
$\{\mathbf{x}_v\}_{v=1}^{|\mathcal{E}|}$. The complexity is computed separately for the real and imaginary components of the pressure field $p_{\text{ref}, i}$:
\begin{equation}
C_{\text{real}, i} = \frac{1}{|\mathcal{E}|-1} \sum_{v=1}^{|\mathcal{E}|-1} \left| \Delta \mathrm{Re}\left(p_{\text{ref}, i}(\mathbf{x}_v)\right) \right|,
\quad
C_{\text{imag}, i} = \frac{1}{|\mathcal{E}|-1} \sum_{v=1}^{|\mathcal{E}|-1} \left| \Delta \mathrm{Im}\left(p_{\text{ref}, i}(\mathbf{x}_v)\right) \right|,
\end{equation}
where $\Delta$ denotes the forward difference operator applied along the ordered sample sequence, defined as $\Delta g(\mathbf{x}_v) \triangleq g(\mathbf{x}_{v+1}) - g(\mathbf{x}_v)$. The sound field complexity at frequency index $i$ is then given by:
\begin{equation}
C_i = \sqrt{C_{\text{real}, i}^2 + C_{\text{imag}, i}^2}.\label{eq:complexity}
\end{equation}

\bmhead{Data availability}
The acoustic simulation and measurement data used in this study are available from the corresponding author for review purposes and will be made publicly available upon acceptance.

\bmhead{Code availability}
The inference code and key scripts are available from the corresponding author for review purposes. A public repository containing the code will be established and linked upon acceptance of the manuscript.

% \bibliography{sn-bibliography}% common bib file

\bmhead{Acknowledgements} 
Y.X. acknowledges partial financial support from the Korea Institute of Machinery and Materials (KIMM). The authors acknowledge the DTU Computing Center for providing computational resources. We also thank Nikolas Borrel-Jensen for his assistance with the code implementation.

\bmhead{Author contributions}
Y.X. designed and constructed the compact microphone array, performed the sensor calibration and experimental measurements, developed the neural field framework, and wrote the original draft. X.L. and M.C. contributed to the design of the neural network architecture and participated in scientific discussions. A.E.K. provided guidance on the mathematical formulation, participated in the interpretation of results, and reviewed and edited the manuscript. C.H.J. supervised the entire project, conceived the research direction, and provided theoretical guidance on acoustics. All authors discussed the results and approved the final manuscript.

\bmhead{Competing interests}
Y.X. and C.H.J. are exploring commercial opportunities related to the technology described in this manuscript.

\end{document}